\documentclass[fleqn,12pt,twoside]{article}
\usepackage{espcrc1}

\usepackage{graphicx}

\newcommand{\AmS}{{\protect\the\textfont2
  A\kern-.1667em\lower.5ex\hbox{M}\kern-.125emS}}
\newcommand{\feoh}{[{\rm Fe}/{\rm H}]}
\newcommand{\nuc}[1]{${}^{#1}$}
\newcommand{\nucm}[2]{{}^{#1}{\rm #2}}
\newcommand{\msun}{M_\odot}
\newcommand{\mmps}{HE0107-5240}

\title{The origin of HE0107-5240 and the production of O and Na in extremely metal-poor stars}

\author{T. Suda\address[DPHU]{Department of Physics, Hokkaido University, \\
        Kita 10 Nishi 8, Kita-ku, Sapporo, Japan}
        \address[VBL]{Meme Media Laboratory, Hokkaido University}%
        \thanks{This work is in part supported by a Grant-in-Aid
        for Science Research from the Japanese Society for the Promotion of Science
        (grant no.15204010).},
        M. Aikawa\addressmark[DPHU]%
        \addressmark[VBL],
        T. Nishimura\addressmark[DPHU],
        M. Y. Fujimoto\addressmark[DPHU],
        and
        I. Iben Jr.\addressmark[DPHU]\thanks{Visiting professor, Hokkaido University, an Eminent Scientist Award, Japanese Society for the Promotion of Science.}%
        \address{Departments of Astronomy and of Physics, University of Illinois}}
       
\begin{document}

\maketitle

\begin{abstract}

We elaborate the binary scenario for the origin of HE0107-5240, the most
metal-poor star yet observed ($\feoh = -5.3$), using current knowledge
of the evolution of extremely metal-poor stars. From the observed C/N value,
we estimate the binary separation and period. Nucleosynthesis in a
helium convective zone into which hydrogen has been injected allows
us to discuss the origin of surface O and Na as well as the abundance
distribution of {\it s}-process elements. We can explain the observed
abundances of $^{12}$C, $^{13}$C, N, O, and Na and predict future
observations to validate the Pop III nature of HE0107-5240. 

\end{abstract}

\section{Introduction}
The existence of a star with \feoh $= -5.3$ has had a great impact on
our understanding of circumstances in the early universe. In spite of its very
small metallicity, \mmps\ shows large enhancements of CNO elements, 
($[\nucm{}{C}/\nucm{}{Fe}] = 4.0$, $[\nucm{}{N}/\nucm{}{Fe}] = 2.3$,
and $\mbox{[O/Fe]}=2.4^{+0.2}_{-0.4}$), a mild enhancement of
Na ($[\nucm{}{Na}/\nucm{}{Fe}] = 2.3$), but no enhancement of main {\it s}-process
elements ($[\nucm{}{Ba}/\nucm{}{Fe}] < 0.8$)\cite{chr02,bes04}.
The observed ratio of C to N ($\sim 150$) cannot be accounted for
by the evolution of an isolated star of mass $M = 0.8 \msun$
\cite{fuj00,pic04,wei04}.

Several attempts to explain the abundance pattern of \mmps\
rely on supernova (SN) nucleosynthesis \cite{ume03,lim03} and the
evolution of a second generation star born from Big Bang matter polluted
by the supernova ejectum \cite{pic04,wei04,sud04}.
The detailed analysis of low mass and extremely metal poor models \cite{pic04,wei04}
are consistent with a case I scenario \cite{fuj00}.

In this paper we describe a first generation binary scenario for
explaining the observed properties of \mmps. We pay particular
attention to neutron capture reactions that play a key role
in systems which consist initially of an $0.8 \msun$ secondary
and a 1-4 $\msun$ primary.

\section{Binary Scenario for HE0107-5240}

\begin{figure}[htb]
\begin{minipage}[t]{150mm}
\includegraphics[width=150mm]{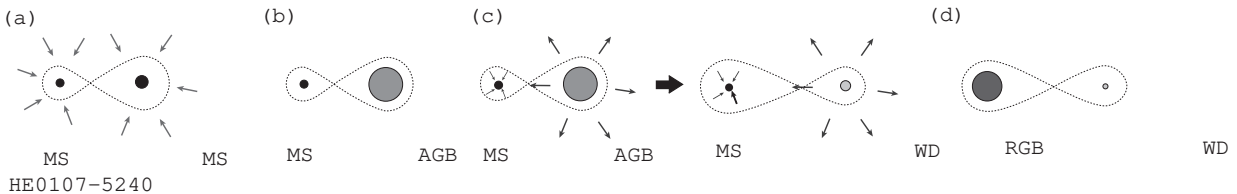}\\
\caption{(a) Wide binary (secondary = $0.8M_{\odot}$ star, primary = $1.2-3M_{\odot}$ star) forms from
primordial cloud. SNe pollute cloud and binary stars acquire surface layers of enhanced Fe by accreting
from polluted cloud.
(b) Primary experiences He-FDDM on the early AGB.
(c) Primary experiences 3rd Dredge-up episodes which produce required abundance ratios in its envelope, and
develops superwind. Some of the wind matter is accreted by the secondary and orbital separation
increases due to angular momentum loss.
(d) MS star evolves into red giant and is observed as HE0107-5240. Secondary has become WD.
}
\label{fig:scenario}
\end{minipage}
\end{figure}

The binary scenario for our Pop.~III model is illustrated in Fig.~\ref{fig:scenario}.
We postulate that only after it has been formed out of primordial matter
does the binary accrete gas from interstellar matter that has been polluted by
the ejecta of Pop.~III SNe (a). After the primordial cloud
disperses, accretion of interstellar matter by the binary as it travels through the Galaxy
is negligible \cite{sud04}. The primary experiences a helium-flash driven deep
mixing (He-FDDM) episode as an EAGB star (b) and products
of the nucleosynthesis in the deep interior are dredged up into surface layers \cite{sud04}.
During the TPAGB phase,
additional carbon is dredged to the surface and the primary emits a wind. The amount
of matter accreted from this wind by the secondary is a function of the assumed orbital
characteristics (c). By requiring that, when the secondary
becomes a red giant, the abundances at its surface match those of \mmps, we estimate
that a total mass of $\sim 0.01 \msun$ is accreted by the secondary and that the primary
experienced of the order of 30 third dredge-up episodes. This translates
into an initial orbital separation of 18 AU and period of 76 yr. Consideration of angular
momentum loss from the binary system suggests final values of 34 AU and 150 yr.
From the amount of mass transferred and the number of dredge-up episodes experienced,
we infer that that the primary emitted mass at the rate ${\dot M}_{\rm loss} \sim 10^{-5}
\msun$/yr, typical of the superwind sustained by ordinary AGB stars.

\section{Progress of He-FDDM and Nucleosynthesis in the helium convective zone}

\begin{figure}[htb]
\begin{minipage}[t]{80mm}
\includegraphics[width=74mm]{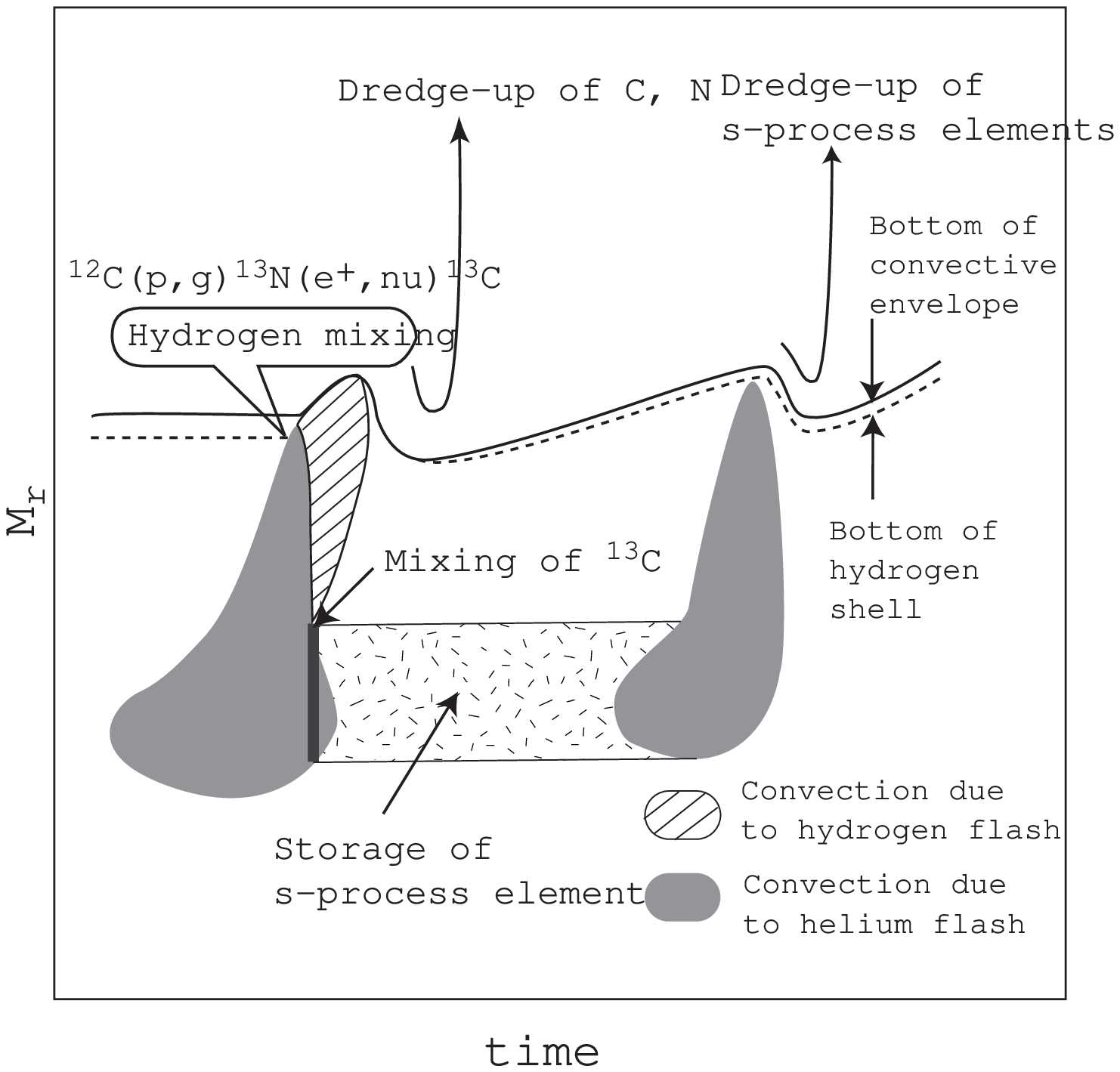}
\caption{
Schematic representation of the time dependence of (1) convective
zones associated with helium-shell flashes and (2) sites for the
nucleosynthesis and/or storage of {\it s}-process elements associated
with these flashes.
He-FDDM begins when hydrogen is ingested by a convective
zone driven by the energy flux from a unique helium-burning flash
in an extremely metal-poor ($\feoh \leq - 2.5$) model star 
of low or intermediate mass at the beginning of the TPAGB phase.
}
\label{fig:convection}
\end{minipage}
\hspace{\fill}
\begin{minipage}[t]{75mm}
\includegraphics[width=74mm]{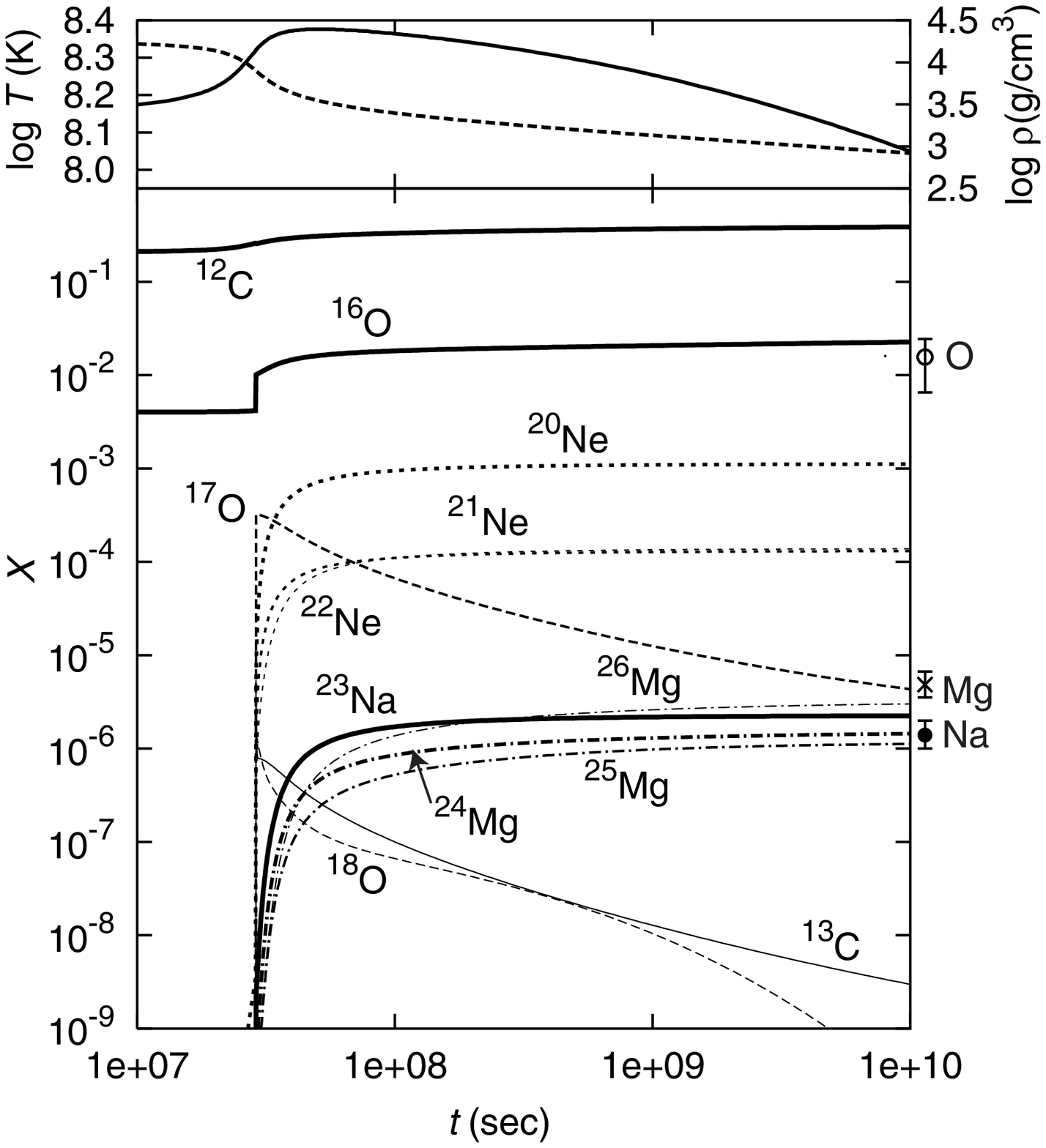}
\caption{
The lower panel shows the variation with time of isotopic
abundances in a helium-flash driven convective zone after
\nuc{13}{C} has been mixed into the zone. 
     The top panel shows the variations in the temperature
(solid line) and in the density (broken line) during the helium
flash.
     Abundances relative to carbon observed for \mmps\
are plotted at the right-hand side of the lower panel for
\nuc{}{O} (open circle), \nuc{}{Na} (filled circle),
and \nuc{}{Mg} (cross).
}
\label{fig:neutron}
\end{minipage}
\end{figure}

The character of mixing during a He-FDDM event is illustrated schematically
in Fig.~\ref{fig:convection}.  
     As the helium shell flash develops, the helium convective zone
grows in mass until, near the peak of the flash, its outer edge
extends into the hydrogen-rich layer.
     Hydrogen is carried downward by convection until
it reaches a point where the lifetime of a proton becomes less
than the convective mixing timescale; then, hydrogen burns
via the $\nucm{12}{C} (p, \gamma) \nucm{13}{N}$ reaction.
     The energy flux due to this reaction causes the convective zone
to split into two parts, the outer
one driven by the energy flux due to hydrogen burning and the
inner one driven by the energy flux due to helium burning.  
     The convective shell engendered by hydrogen burning transports
C- and N-rich matter outwards where it can ultimately be dredged up and incorporated
into the surface convective zone.
     In models with large core masses, the convective shell
sustained by helium burning persists even after the hydrogen-driven
convective zone has disappeared \cite{fuj00,iwa04}.

Before the split into two convective zones, some of the mixed-in
hydrogen is captured by \nuc{12}C; the product is carried inward and,
after the split, is trapped in the surviving helium convective
zone as \nuc{13}{N} and/or as its daughter \nuc{13}C.  
      At the high temperatures in the surviving helium convective shell,
the reaction $\nucm{13}{C} (\alpha, n) \nucm{16}{O}$ occurs, with
interesting consequences, as demonstrated by \cite{iwa04} for a
$2 \msun$ star with $\feoh = -2.7$. 

To explore the ensuing neutron-capture nucleosynthesis in the
helium convective shell, we adopt the same one-zone approximation
used by \cite{aik01}.  
     We treat the amount of mixed \nuc{13}C as a parameter
since this amount varies with the strength of the helium shell flash,
which depends on stellar mass, and, also, is a function of the treatment
of convection. The mixing is assumed to occur at the peak of the shell
flash.  

Fig.~\ref{fig:neutron} illustrates the progress of nucleosynthesis in the
helium convective zone in a Pop.~III model star,
when the amount of \nuc{13}{C} mixed into this
zone is chosen in such a way that
$\nucm{13}{C}/\nucm{12}{C} = 10^{-3}$.
     As soon as $\nucm{13}{C}$ is mixed into the convective zone, it
rapidly reacts with helium to produce neutrons.   
     The neutrons so produced are captured primarily by $\nucm{12}{C}$
to form $\nucm{13}{C}$ and then reappear in consequence of additional
$\nucm{13}C (\alpha, n) \nucm{16}{O}$ reactions. This neutron
cycling process continues until the $\nucm{16}O (n,\gamma)
\nucm{17}O$ reaction has converted most of the initially injected
$\nucm{13}C$ into \nuc{17}O.  
     Then, $\alpha$ capture on \nuc{17}{O}, which occurs $\sim 10^{4}$
times more slowly than capture on \nuc{13}{C} at the relevant
temperatures, starts to produce neutrons via the reaction
$\nucm{17}O (\alpha, n) \nucm{20}{Ne}$.
     The newly formed \nuc{20}{Ne} consumes neutrons
to yield heavier isotopes of Ne and isotopes of Na and Mg.
If the iron-group elements exist in the helium convective zone, most of them
are converted into heavy {\it s}-process elements.
Therefore, in the case of stars formed from a polluted cloud or stars with helium
convective zones that were polluted by accreted metals during He-FDDM,
the upper limit of the surface lead enhancement for \mmps\ becomes [Pb/Fe] $= 1-2$.
     We emphasize that the appearance
of \nuc{13}{C} in a convective shell occurs only once. During subsequent
thermal pulses, any \nuc{13}{C} that might be formed because of extra-mixing
would be burned radiatively between pulses \cite{str95}. However, the
observed sharp decrease in the [Pb/Ba] ratio for [Fe/H] $< -2.5$ suggests
that this mechanism does not operate in extremely metal-poor stars.

\section{Conclusions}
We can explain C, N, O, Na, and $^{12}$C/$^{13}$C quantitatively by the binary scenario.
The scenario predicts a current orbital period of $\sim$ 150 yr and continuous observations
over an extended period ($>$10 years) may reveal the predicted variations in radial velocity.
It also predicts the enhancement of $^{25}$Mg and $^{26}$Mg relative to $^{24}$Mg.
A deficiency of Pb by $[{\rm Pb}/{\rm H}] \sim -3$ is needed to establish conclusively
that \mmps\ is a Pop.III star.

\end{document}